\documentstyle[titlepage,12pt]{article}

\def\be{\begin{equation}}
\def\tphi{\tilde{\phi}}
\def\tpsi{\tilde{\psi}}
\def\grad{\nabla}
\def\ee{\end{equation}}

\newcount\sectionnumber
\def\sect
{\global\equationnumber=0
\global\advance\sectionnumber by 1
\the\sectionnumber . }
\newcount\equationnumber
\def   \num
{\eqno{\global\advance\equationnumber by 1
\left(\the\sectionnumber .\the\equationnumber \right)}}

\begin{document}
\title{Two Dimensional Quantum Gravity Coupled to Matter}
\author{R.B. Mann \\
Department of Physics \\
University of Waterloo \\
Waterloo, Ontario \\
N2L 3G1}

\date{June 24, 1992 \\
WATPHYS TH-90/04}

\maketitle

\begin{abstract}

A classical two dimensional theory of gravity which has a number of
interesting features (including a Newtonian limit, black holes and
gravitational collapse) is quantized using conformal field theoretic
techniques. The critical dimension depends upon Newton's
constant, permitting models with $d=4$. The constraint
algebra and scaling properties of the model are computed.

\end{abstract}

\section{Introduction}

It has become clear in recent years that two dimensional quantum gravity
has much to teach us, both in terms of obtaining a proper understanding of
two dimensional systems and as a theoretical laboratory for
the $(3+1)$ dimensional case. In the latter context it is hoped that the
reduction in the number of degrees of freedom  will yield a better
understanding  of short-distance problems, topology change, singularities
and the  cosmological constant problem. In the former context the problem
of confronting two dimensional quantum gravity  was avoided by by
considering massless matter systems. Classically, general coordinate
invariance and Weyl invariance may be used to gauge fix all three degrees
of freedom of the metric \cite{Polyakov1};  upon quantization, Weyl
invariance no longer holds except for particular values of the central
charge of the matter ({\it i.e.} the critical dimensions).  By restricting
attention to these special cases the problem of quantizing the two
dimensional gravitational field itself was avoided.  This restriction was
lifted when it was shown  in light cone gauge that that such systems can be
quantized in the  presence of the Liouville mode of the metric for a genus
zero surface \cite{KPZ}. Shortly afterward it was shown how to obtain this
result in conformal gauge for all genera \cite{DDK}.  However, the central
charge in such models is restricted to be less than or equal to one, and it
is not clear how to recover the known critical cases.

A key problematic element in the above program is the choice of classical
action for the gravitational field.   The Einstein tensor  vanishes in two
dimensions for all metrics and the Einstein-Hilbert action simply yields
the Euler number of the manifold.  A number of suggestions have been made
to address this problem
\cite{D'Hoker,Jackiw,Teit,WitCMP,LPW,CWIT,Polch}. These
typically involve taking the classical action to be trivial (either zero or
the Einstein Hilbert action) and then gauge-fixing the symmetries of the
metric to obtain a quantum action, or alternatively making use of
topological interactions \cite{CWIT}, a variant of which
\cite{Cham} permits constant curvature solutions classically.
In all cases the matter-gravity interaction is classically trivial, and
quantum-mechanically quite limited.

However in the last few years it has been shown that classical gravity in
two spacetime dimensions need not be so trivial \cite{MFound}. An
interesting relativistic theory of gravitation in this context may be
formed by setting the Ricci scalar $R$ equal to $T=T_\mu^{\ \mu}$, the
trace of the conserved stress energy tensor
\be
R = 8\pi G T   \label{1}
\ee
where $G$ is Newton's constant in two spacetime dimensions.
In spite of its simplicity, this theory has a number of remarkable
classical and semi-classical features, including a well-defined Newtonian
limit \cite{MFound}, black holes \cite{MST,DanRobb},
a post-Newtonian expansion,
gravitational waves, FRW cosmologies, gravitational collapse \cite{Arnold}
and black hole radiation \cite{TomRobb,2dsemi,Shardir}.   These features
suggest that it is potentially a very useful tool in the study of quantum
gravity: since its classical features are so similar to those of $(3+1)$
dimensional general relativity, one might hope that its quantization would
bear a similar resemblance to $(3+1)$ dimensional quantum gravity. Indeed,
it has been shown the system (\ref{1}) along with the conservation of
stress-energy
\be
\nabla_\nu T^{\mu\nu} = 0               \label{2}
\ee
may be understood as the $D\to 1$ limit of  the $(D+1)$ dimensional
Einstein equations \cite{RobbGRG}. The present paper is concerned with
taking a first step towards quantization of this theory.

The system (\ref{1},\ref{2}) may be derived from the action  \cite{2dsemi}
\begin{equation}
S=\frac{1}{8\pi G}\int d^2x\,\sqrt{-g}\left(\frac{1}{2}g^{\mu\nu}
\partial_\mu\psi\partial_\nu\psi + \psi R
- 8 \pi G {\cal L}_M \right)\label{3}
\end{equation}
which in turn yields the field equations
\begin{equation}
\frac{1}{\sqrt{-g}}\partial_\mu\left[\sqrt{-g}g^{\mu\nu}\partial_\nu
\psi\right] - R = 0
\label{4}
\end{equation}
\begin{equation}
\frac{1}{2}\left(\psi_{;\mu}\psi_{;\nu} - \frac{1}{2} g_{\mu\nu}
\psi_{;\alpha}\psi_{;\alpha}\right) +
g_{\mu\nu}g^{\alpha\beta}\psi_{;\alpha;\beta} - \psi_{;\mu;\nu}
= 8\pi G T_{\mu\nu} \label{5}
\end{equation}
plus the equations of motion for the matter fields.  Insertion of the trace
of (\ref{5}) into (\ref{4}) yields (\ref{1}), and conservation of the
stress energy tensor (eq. (\ref{2})) is guaranteed since the covariant
divergence of the left hand side of (\ref{5}) is identically zero when
(\ref{4}) holds. This also follows from considering the invariance of the
action (\ref{3}) under infinitesimal co-ordinate transformations, analogous
to the $(3+1)$-dimensional case.

The action (\ref{3}) with $T_{\mu\nu}=0$ has been considered before
\cite{MarnTor,HMS} in the context of finding a local action which produces
the Polyakov theory. In such cases the $\psi$-field is fully eliminated in
terms of the metric degree of freedom by imposing boundary conditions which
break reparametrization invariance, thereby inducing an anomaly into the
Virasoro algebra. In the present case,  $\psi$ is treated as an auxiliary
field, since the classical evolution of the gravity/matter system is
independent of the evolution of $\psi$. However the $\psi$ field obeys the
equation
\be
\frac{1}{2}\left(\psi_{;\mu}\psi_{;\nu} - \frac{1}{2} g_{\mu\nu}
\psi_{;\alpha}\psi_{;\alpha}\right) + \frac{1}{2}
g_{\mu\nu}g^{\alpha\beta}\psi_{;\alpha;\beta} - \psi_{;\mu;\nu} = 8\pi G
(T_{\mu\nu}-\frac{1}{2}g_{\mu\nu}T)  \label{6}
\end{equation}
and so depends upon the evolution of the gravity/matter system
(\ref{1},\ref{2}).

Equations (\ref{4}) and (\ref{5})  form a system of 4 equations with 2
identities, which is equal to the number of unknowns (the metric degree of
freedom and $\psi$). Reorganized in the form (\ref{1},\ref{6}) (and
(\ref{2})), one easily sees that upon  setting
$T_{\mu\nu}=g_{\mu\nu}\Lambda$ that they reduce to a  proposal considered
previously by Jackiw and Teitelboim for two dimensional gravity
\cite{Jackiw,Teit}, whose quantization was recently considered by
Chamseddine \cite{Cham}.  Such a restriction permits only constant curvature
solutions to the field equations.  In contrast to this the theory based on
(\ref{1}) and (\ref{6}) allows curvature and stress-energy to act upon one
another in a manner which is very similar to the $(3+1)$ dimensional
situation.

Motivated by the above, the classical gravitational action is taken to be
(\ref{3}), and its quantum properties are considered herein.  The Newtonian
gravitational constant $G$ will be seen to significantly modify the
relationship between the central charge and the conformal dimensions of the
matter fields, permitting critical dimensions larger than 1.

\section{Constraint Algebra}

It is instructive to consider the constraint algebra which follows from
(\ref{3}). Working in conformal gauge, with $g_{\mu\nu}=e^\phi
\eta_{\mu\nu}$ yields from (\ref{4}) and (\ref{5})
\begin{eqnarray}
\frac{1}{4}\left(\dot{\psi}^2+(\psi^{\prime})^2\right) - \psi^{\prime\prime}
+\frac{1}{2}\dot{\phi}\dot{\psi} + \frac{1}{2}\phi^{\prime}\psi^{\prime}
&=&T_{00} \label{7a} \\
\frac{1}{2}\dot{\psi}\psi^{\prime} + \frac{1}{2}\phi^{\prime}\dot{\psi}
+ \frac{1}{2}\psi^{\prime}\dot{\phi} - \dot{\psi}^{\prime} &=& T_{01}
\label{7b}
\end{eqnarray}
for the constraint equations and
\begin{eqnarray}
\frac{1}{4}\left(\dot{\psi}^2+(\psi^{\prime})^2\right) - \ddot{\psi}
+\frac{1}{2}\dot{\phi}\dot{\psi} + \frac{1}{2}\phi^{\prime}\psi^{\prime}
&=&T_{11} \label{8a} \\
\ddot{\psi} - \psi^{\prime\prime} + \ddot{\phi} - \phi^{\prime\prime}
&=& 0 \label{8b}
\end{eqnarray}
for the dynamical equations.  Here $\psi^\prime
\equiv \partial\psi/\partial x$ and
$\dot{\psi} \equiv \partial\psi/\partial t$.
Subtracting (\ref{8a}) from (\ref{7a}) allows
this latter set to be rewritten as
\begin{eqnarray}
\ddot{\phi} - \phi^{\prime\prime} &=& -\left(T_{00}-T_{11}\right) \label{9a}
\\
\ddot{\psi} - \psi^{\prime\prime} + \ddot{\phi} - \phi^{\prime\prime}
&=& 0 . \label{9b}
\end{eqnarray}

Following ref. \cite{MarnTor}, it is straightforward to show that
\be
P_\psi = -\left(\dot{\psi} + \dot{\phi}\right)  \label{10}
\ee
is the momentum conjugate to $\psi$ and that the momenta conjugate to the
metric  is
\be
\Pi^{00} = 0 \quad \Pi^{01}=-\frac{1}{2}e^{-\phi}\psi^{\prime} \quad
\Pi^{11} = e^{-\phi}\dot{\psi}   \label{11}
\ee
where $\Pi^{\alpha\beta}\equiv \frac{\delta S}{\delta g^{\alpha\beta}}$.
Clearly only $P_\psi$ and $\Pi^{11}$ are independent canonical variables in
conformal gauge.  It is useful to make a canonical transformation on these
variables so that
\be
\chi = \psi + \phi \qquad P_\chi = P_\psi = -\dot{\chi} \label{12}
\ee
and
\be
\Pi = e^\phi \Pi^{11} = \dot{\phi}  \qquad .
\ee
This guarantees that on a spatial slice
\begin{eqnarray}
\left\{ \chi(x), P_\chi(x')\right\} &=& \delta(x-x')  \label{13a} \\
\left\{ \psi(x), \Pi(x')\right\} &=& \delta(x-x')  \label{13b}
\end{eqnarray}
are the canonical Poisson brackets.

Under this transformation the constraint equations (\ref{7a}) and
(\ref{7b}) become respectively
\be
a^2_{\chi+} - b_{\chi+} - (a^2_{\phi-} + b_{\phi-}) +
a^2_{\chi-} + b_{\chi-} - (a^2_{\phi+} - b_{\phi+}) = 2 T_{00} \label{14}
\ee
\be
-(a^2_{\chi+} - b_{\chi+}) + (a^2_{\phi-} + b_{\phi-}) +
a^2_{\chi-} + b_{\chi-} - (a^2_{\phi+} - b_{\phi+}) = 2 T_{01} \label{15}
\ee
where
\be
a_{\chi\pm} \equiv \frac{1}{2}\left(P_\chi \pm \chi'\right) \quad
a_{\phi\pm} \equiv \frac{1}{2}\left(P_\phi \pm \phi'\right) \label{16}
\ee
and
\be
b_{\chi\pm} \equiv 2a^\prime_{\chi\pm} \quad
b_{\phi\pm} \equiv 2a^\prime_{\phi\pm} \quad .
\label{17}
\ee
On spatial sections topologically equivalent to $S^1$ these variables may
be Fourier transformed so that
\begin{eqnarray}
L^m_{\chi\pm} &=& \int_{-\pi}^\pi d\sigma e^{-im\sigma}
(a^2_{\chi\pm} \mp b_{\chi\pm})   \label{18a} \\
L^m_{\phi\pm} &=& \int_{-\pi}^\pi d\sigma e^{-im\sigma}
(a^2_{\phi\pm} \mp b_{\phi\pm})   \label{18b}
\end{eqnarray}
where the periodic spatial variable $\sigma\in (-\pi,\pi)$. The Poisson
brackets of these variables are
\be
\left\{L^n_{\chi\pm} , L^m_{\chi\pm}\right\} =
\pm i (n-m) L^{n+m}_{\chi\pm} \mp 4\pi i n^3 \delta_{n+m,0} \label{19}
\ee
\be
\left\{L^n_{\phi\pm} , L^m_{\phi\pm}\right\} =
\pm i (n-m) L^{n+m}_{\phi\pm} \mp 4\pi i n^3 \delta_{n+m,0} \label{20}
\ee
all other brackets being zero.

Adding the constraints (\ref{14}) and (\ref{15}) yields respectively
\be
{\cal L}^m_{\pm} = \left(\tilde{T}^m_{00} \mp \tilde{T}^m_{01}\right)
\label{21}
\ee
where  $\tilde{T}^m_{\mu\nu}$ is the Fourier transform of the stress-energy
tensor. The quantities
\be
{\cal L}^m_{\pm} \equiv L^m_{\chi\pm} - L^m_{\phi\mp}  \label{22}
\ee
obey
\be
\left\{{\cal L}^n_{\pm} , {\cal L}^m_{\pm}\right\} =
i (n-m) {\cal L}^{n+m}_{\pm}  \label{23}
\ee
for their Poisson bracket algebra.

Each of the $\chi$, gravity sectors yields a classical Virasoro algebra with
a non-zero central charge.  However these cancel out in
the {\it combined} $\chi$-gravity system, as is clear from
(\ref{22},\ref{23}).  Classical coupling to conformally invariant
matter will not affect this
result, since the stress-energy tensor will in general have a Fourier
decomposition in terms of operators which obey the Virasoro
algebra (\ref{23}). For example, a massless scalar field $\varphi$ will
have stress-energy tensor components
\be
{T}_{00} \pm {T}_{01} =
\frac{1}{2}(P_\varphi \pm \varphi')^2   \label{24}
\ee
which upon Fourier decomposition will yield operators $L^m_{\varphi\pm}$
whose Poisson brackets form a Virasoro algebra with vanishing central
charge.

Other approaches to two-dimensional gravity typically impose additional
constraints to the system described by the action (\ref{3}) which introduce
an anomaly at the classical level. For example, if the $\psi$ degree of
freedom is frozen out ({\it i.e.} if in (\ref{8b}) there are no homogenous
solutions permitted for $\psi$) then $\chi=0$ and consequently $P_\chi=0$.
The action (\ref{3}) then becomes the non-local Polyakov action
\cite{Polyakov1} and the constraint algebra develops an anomaly (\ref{23})
due to this non-locality. Conformally coupling to two-dimensional matter
and quantizing then permits one to cancel off this anomaly against the
central charge of the matter \cite{Polch}. Alternatively, one could fix the
two-dimensional metric $g_{\mu\nu}=\hat{g}_{\mu\nu}$ which would freeze out
the $\phi$ degree of freedom. In this case the action (\ref{3}) becomes
that of the Liouville action \cite{Seiberg}.  The field $\psi$ is
interpreted as the quantum-dynamical field whose exponential yields a
conformal factor which provides the quantum corrections to
$\hat{g}_{\mu\nu}$ {\it i.e.} the full two-dimensional metric
$g_{\mu\nu}=e^\psi\hat{g}_{\mu\nu}$  \cite{KPZ,DDK}.

The approach taken here does not involve freezing out any of the degrees of
freedom in the system described by the action (\ref{3}).
Classically the metric is locally $g_{\mu\nu}=e^\phi \eta_{\mu\nu}$ and the
$\psi$ field is an auxiliary field whose classical evolution has no effect
on the gravity/matter system.

\section{Quantum Corrections}

Following the remarks of the previous section, quantization of two
dimensional gravity coupled to matter may be carried out by considering
a functional integral over the field configurations of the metric, $\psi$,
a and matter fields. This may be done using the path integral
\be
Z=\int \frac{{\cal D}g}{V_{GC}}
{\cal D}\psi {\cal D}\Phi e^{-S[\psi,g] -S_M[\Phi]}  \label{25}
\ee
where $S=S[\psi,g]+S_M[\Phi]$ is the Euclideanization of the action
(\ref{3}), with $S_M$ being the matter part of the action and $\Phi$
representing the matter fields. The volume of the diffeomorphism group,
$V_{GC}$, has been factored out.

Making the same scaling assumption as in refs. \cite{DDK,Cham} about
the functional measure  yields
\begin{eqnarray}
Z&=&\int [{\cal D}\tau] {\cal D}_g\phi {{\cal D}_g b} {{\cal D}_g c}
{\cal D}_g\psi {\cal D}_g\Phi e^{-S[\psi,\phi]-S_{gh}[b,c]-S_M[\Phi]}
\nonumber  \\
&=&\int [{\cal D}\tau] {\cal D}_{\hat{g}}\phi {{\cal D}_{\hat{g}} b}
{{\cal D}_{\hat{g}} c} {\cal D}_{\hat{g}}\psi {\cal D}_{\hat{g}}\Phi
e^{-S[\psi,\phi]-S_{gh}[b,c]-S_M[\Phi]-\hat{S}[\phi,\hat{g}]}
\label{26}
\end{eqnarray}
where $[{\cal D}\tau]$ represent the integration over the Teichmuller
parameters and
\be
\hat{S}[\phi,\hat{g}]=\frac{1}{8\pi}\int d^2x \sqrt{\hat{g}}\left(
\hat{g}^{\mu\nu}\partial_\mu\phi\partial_\nu\phi - Q \phi\hat{R}
+\mu e^{\alpha\phi}\right)
\label{27}
\ee
is the Liouville action with arbitrary coefficients and $g_{\mu\nu}=
e^{\alpha\phi}\hat{g}_{\mu\nu}$. Note that $R[e^{\alpha\phi}\hat{g}] = e^{-
\alpha\phi}(\hat{R} -\alpha\hat{\grad}^2\phi)$ where $\hat{R}$  and
$\hat{\grad}^2$  are respectively the curvature scalar and Laplacian of the
metric  $\hat{g}$. $S_{gh}$ is the action for the ghost fields $b$ and $c$.

The approach is then to determine the parameters $\alpha$ and $Q$ from the
requirement that the conformal anomaly vanish and that $e^{\alpha\phi}$ is
a conformal tensor of weight (1,1). The action $S_{\rm TOT} =
\hat{S}+S[\psi,g]+S_{gh}+S_M$ may be rewritten as
\begin{eqnarray}
{S}_{\rm TOT}&=&  S_{gh} + S_M +
\frac{1}{8\pi}\int d^2x \sqrt{\hat{g}}\left[
(1+\frac{\alpha^2}{2G})\hat{g}^{\mu\nu}\partial_\mu\phi\partial_\nu\phi
- (Q-\frac{\alpha}{G})\phi\hat{R} \right. \nonumber \\
&& -\frac{1}{G}\left(
\frac{1}{2}\hat{g}^{\mu\nu}\partial_\mu(\psi+\alpha\phi)
\partial_\nu(\psi+\alpha\phi) (\psi+\alpha\phi)\hat{R}\right)\nonumber \\
&& +(\mu+8\pi\Lambda)e^{\alpha\phi}
\bigg]  \label{27a}
\end{eqnarray}
where $\Lambda$ is the (possibly zero) classical cosmological constant.
Upon rescaling the fields $\psi$ and $\phi$ so that
\be
\tilde{\phi}=\sqrt{1+\frac{\alpha^2}{2G}}\phi \quad
\tilde{\psi}=\frac{1}{\sqrt{2G}}\left(\psi+\alpha\phi\right)
\label{28}
\ee
(\ref{27a}) becomes
\begin{eqnarray}
S_{\rm TOT}&=& \frac{1}{8\pi}\int d^2x \sqrt{\hat{g}}\left\{
\hat{g}^{\mu\nu}\partial_\mu\tilde{\phi}\partial_\nu\tilde{\phi}
- \frac{Q-\frac{\alpha}{G}}{\sqrt{1+\frac{\alpha^2}{2G}}}\phi\hat{R}
-\hat{g}^{\mu\nu}\partial_\mu\tilde{\psi}
\partial_\nu\tilde{\psi} \right.  \nonumber \\
&&\left. - \sqrt{\frac{2}{G}}\tilde{\psi}\hat{R}
+(\mu+8\pi\Lambda)
\exp\left[\frac{\alpha}{\sqrt{1+\frac{\alpha^2}{2G}}}\tphi\right]
\right\} + S_{gh} + S_M
\quad . \label{29}
\end{eqnarray}
The coefficients in front of the $\tilde{\psi}$ terms are due to
the sign of the kinetic energy term in (\ref{3}). This may be
dealt with by rescaling $\tpsi\to i\tpsi$ so that the functional
integral converges.

The action in (\ref{29}) may now be analyzed using conformal field
theoretic techniques.  The fields $\tphi$, $\tpsi$ have propagators
\be
<\tphi(z)\tphi(w)> = - \ln(z-w) = <\tpsi(z)\tpsi(w)>  \label{30}
\ee
and the contribution of the matter and ghost fields is as usual.
Hence it is straightforward to compute the total central charge
\begin{eqnarray}
c_{\rm TOT} &=& 1 + 3\frac{(Q-\frac{\alpha}{G})^2}{1+\frac{\alpha^2}{2G}} + 1
-\frac{6}{G} -26 + c_M \nonumber \\
&=& c_M-24 + 3\frac{GQ^2-2(1+\alpha Q)}{G+\alpha^2/2}
\label{31}
\end{eqnarray}
using the operator product expansion of the stress-energy tensor which is
associated with (\ref{29}).  The anomalous dimension of the (renormalized)
cosmological constant term in (\ref{29}) is
\be
\Delta = -\frac{1}{2}\frac{\alpha}{\sqrt{1+\frac{\alpha^2}{2G}}}\left(
\frac{Q-\alpha/G}{\sqrt{1+\frac{\alpha^2}{2G}}}+
\frac{\alpha}{\sqrt{1+\frac{\alpha^2}{2G}}}\right)   \quad .
\label{32}
\ee
In the $G\to\infty$ limit these results yield what one would obtain from
the Polyakov approach \cite{KPZ,DDK}, except that $c_M\to c_M+1$
due to the presence of the additional scalar $\tpsi$.

Setting $\Delta=1$  so that $e^{\alpha\phi}$ is a conformal tensor of
weight (1,1) yields
\be
\alpha^2+Q\alpha+2=0 \label{33}
\ee
exactly as in ref. \cite{DDK}. This yields the constraint $Q^2>8$ since
$\alpha$ must be real. Setting $c_{\rm TOT}=0$ and using (\ref{33}) yields
\be
\alpha^2 = \frac{(12-c_M)G-6\pm\sqrt{(c_MG-6)((c_M-24)G-6)}}{c_M-12+6G}
\label{34}
\ee
for $\alpha$ in terms of $c_M$.

As $\alpha$ is real, $\alpha^2$ must be positive, imposing constraints on
$c_M$ and $G$. There are three distinct cases to consider.

\noindent
({\bf i}) $c_M<12-6G\quad$ In this case the negative sign for the square
root must be chosen in (\ref{34}) and so
$$
\alpha^2 = \frac{\sqrt{(c_MG-6)((c_M-24)G-6)}+(c_M-12)G+6}{12-6G-c_M}
$$
is the only possible solution.  Positivity of $c_M$ implies $G<2$.

\noindent
({\bf ii}) $c_M>12-6G\quad$ The discriminant in this case is not positive
if $6/G<c_M<24+6/G$.  It is straightforward to show that if $c_M$ is larger
than $24+6/G$ then $\alpha^2$ is not positive, yielding $c_M<6/G$.
There is no upper limit on $G$, but positivity of $\alpha^2$ implies
$G>1$ regardless of which sign of the square root is taken.
In the $G\to\infty$ limit this
is the familiar restriction  $1<c_M+1<25$ of string theory, since
the additional scalar $\psi$ adds one unit of central charge to $c_M$.

\noindent
({\bf iii}) $c_M=12-6G\quad$ In this case
$$
\alpha^2 = 2 \frac{G}{G^2-1}
$$
which implies $2>G>1$ when $c_M$ is positive.

Finally, one can compute the scaling dependence of the correlation
functions. Inserting the factor
\be
1=\int{\cal D}A\delta\left(\int d^2x \sqrt{\hat{g}}e^{\alpha\phi}-A\right)
\label{36}
\ee
into the functional integral (\ref{26}) implies that
\be
<\Pi_i e^{\alpha_i\Phi_i}>_A = A^{\xi-1}
<\Pi_i e^{\alpha_i\Phi_i}>_{A=1}  \label{37}
\ee
with
\be
\xi = \frac{QG-\alpha}{\alpha G}(h-1)  \label{38}
\ee
where $A$ is the area of the surface and $h$ the number of handles.

\section{Discussion}

The model presented here is that of a two-dimensional quantum theory of
gravity which has a non-trivial classical limit.  As such it has a number
of interesting features which merit further study.  Unlike virtually all
other two-dimensional theories of gravity, it has a well-defined Newtonian
limit and its classical features closely resemble those of $(3+1)$
dimensional general relativity \cite{RobbGRG}. The restriction on the
central charge now depends on the dimensionless Newton constant $G$, and so
for small enough $G$ one can avoid the constraints imposed by the critical
dimensions present in other models\footnote{Although the model in ref.
\cite{Cham} has no restriction on the critical dimension, it is also
limited to constant curvature solutions.}
\cite{DDK,Polch}. For example, one could use this model as the basis of a
non-critical string theory in four dimensions provided $G<4/3$.

The restriction to coupling to conformally invariant matter may be lifted
by extending the matter action so that
\be
\tilde{S}_M = S_M + \sum_i m_i {\cal O}_i  \label{39}
\ee
where $m_i$ are (dimensional) constants and ${\cal O}_i$ are operators of
conformal dimension $\Delta_i$. Upon coupling to gravity, this term is
modified to
\be
\tilde{S}_M = S_M + \sum_i m_i e^{\alpha_i \phi}{\cal O}_i \label{40}
\ee
in (\ref{27a}).  Expanding in the parameters $m_i$, one must as before
require that each term in (\ref{40}) be a (1,1) operator and so (\ref{33})
becomes
\be
 -\frac{1}{2}\frac{\alpha_i}{\sqrt{1+\frac{\alpha^2}{2G}}}\left(
\frac{Q-\alpha/G}{\sqrt{1+\frac{\alpha^2}{2G}}}+
\frac{\alpha_i}{\sqrt{1+\frac{\alpha^2}{2G}}}\right) + \Delta_i = 1
\label{41}
\ee
which forms a constraint on the exponents $\alpha_i$.  If $\mu+8\pi\Lambda$
is set to zero in (\ref{27a}), then the restriction that $Q^2>8$ no longer
holds, but instead is modified to
\be
Q^2 > 8(1-\Delta)(1-\frac{\Delta}{G}) \label{42}
\ee
provided at least one of the constants ($m_1=m$, say, with corresponding
operator of dimension $\Delta$) is non-zero.

It is of course possible to extend these results to include negative values
of $G$; although such models have antigravity, they have been of some
field-theoretic interest \cite{MST,BHT}. From (\ref{27a}) it is clear that
in this case a critical value for $G$ exists ($G_{\rm crit}=2\alpha^2$)
beyond which the model is no longer well-defined. At this point the kinetic
term for $\phi$  vanishes and the model becomes singular, suggesting a
phase transition to another model.

\section*{Acknowledgements}

This work was supported by the Natural Sciences and Engineering Research
Council of Canada.

 \par\vfill\eject


\begin{thebibliography}{References}

\bibitem{Polyakov1}A.M. Polyakov, Phys. Lett. {\bf B103} (1981) 207.
\bibitem{KPZ}A.M. Polyakov, Mod. Phys. Lett. {\bf A2} (1987) 899;
V. Kniznik, A.M. Polyakov, A.B. Zamolodchikov  Mod. Phys. Lett.
{\bf A3} (1988) 819.
\bibitem{DDK}F. David, Mod. Phys. Lett. {\bf A3} (1988) 1651; J. Distler
and H. Kawai, Nucl. Phys. {\bf B321} (1989) 509.
\bibitem{D'Hoker}E. D'Hoker and R. Jackiw, Phys. Rev. {\bf D26} (1982)
3517.
\bibitem{Jackiw}C. Teitelboim in {\sl Quantum Theory of Gravity},  ed.
S. Christensen (Adam Hilger, Bristol 1984), p.327; R. Jackiw, {\it ibid}.,
p.~403;  Nucl. Phys. {\bf B252} (1985) 343.
\bibitem{Teit}C. Teitelboim, Phys. Lett. {\bf B126} (1983) 41, 46;
M. Henneaux, Phys. Rev. Lett. {\bf54} (1985) 959; T. Fukuyama and
K. Kimimura, Phys. Lett. {\bf B160} (1985) 259.
\bibitem{WitCMP}E. Witten, Comm. Math. Phys. {\bf 117} (1988) 353.
\bibitem{LPW}J. Labastida, M. Pernici and E. Witten, Nucl. Phys. {\bf B310}
(1988) 611.
\bibitem{CWIT}A.M. Chamseddine and D. Wyler, Phys. Lett. {\bf B228} (1989)
75; Nucl. Phys. {\bf B340} (1990) 545; K. Isler and C. Trugenberger, Phys.
Rev. Lett. {\bf 63} (1989) 834.
\bibitem{Polch}J. Polchinski, Nucl. Phys. {\bf B324} (1989) 123.
\bibitem{Cham}A.M. Chamseddine, Phys. Lett {\bf B256} (1991) 379.
\bibitem{MFound}R.B. Mann, Found. Phys. Lett. {\bf 4} (1991) 425.
\bibitem{MST}R.B. Mann, A. Shiekh, and L. Tarasov, Nucl. Phys. {\bf B341}
(1990) 134.
\bibitem{DanRobb}J.D. Christensen and R.B. Mann, Class. Quant. Grav.
(to be published).
\bibitem{Arnold}A.E. Sikkema and R.B. Mann, Class. Quantum Grav.
{\bf 8} (1991) 219.
\bibitem{TomRobb}R.B. Mann and T.G. Steele, Class. Quant. Grav. {\bf 9}
(1992) 475.
\bibitem{2dsemi}R.B. Mann, S. Morsink, A.E. Sikkema and T.G. Steele,
Phys. Rev. {\bf D43} (1991) 3948.
\bibitem{Shardir}S.M. Morsink and R.B. Mann, Class. Quant. Grav.
{\bf 8} (1991) 2257.
\bibitem{RobbGRG}R.B. Mann, Gen. Rel. Grav. {\bf 24} (1992) 433.
\bibitem{MarnTor}C.G. Torre, Phys. Rev. {\bf D40} (1989)
2588; R. Marnelius, Nucl. Phys. {\bf B211} (1983) 14.
\bibitem{HMS}J.A. Helayel-Nyeto, S. Mokhtari and A.W. Smith,
Phys. Lett. {\bf B236} (1990) 12.
\bibitem{Seiberg}N. Seiberg, Prog. Theo. Phys. Supp. {\bf 102} (1990) 319.
\bibitem{BHT}J.D. Brown, M. Henneaux and C. Teitelboim,
Phys. Rev. {\bf D33} (1986) 319; J.D. Brown, {\sl Lower
Dimensional Gravity}, (World Scientific, 1988).


\end{thebibliography}
\end{document}